\documentclass[12pt]{article}
\usepackage{amsfonts,amssymb}

\advance\hoffset by -1.1truecm
\advance\voffset by -2.0truecm

\newcommand{\be}{\begin{equation}}
\newcommand{\ee}{\end{equation}}

\newcommand{\bea}{\begin{eqnarray}}
\newcommand{\eea}{\end{eqnarray}}
\newcommand{\bean}{\begin{eqnarray*}}
\newcommand{\eean}{\end{eqnarray*}}

\makeatletter
\@ifundefined{newsymbol}
{

\newcommand{\C}{\mbox{C \hspace{-1.16em} \raisebox{-0.018em}{\sf l}}\;}

\newcommand{\I}{\mbox{\rm I}\hspace{-0.5em}\mbox{\rm I}\,}

\newcommand{\R}{\mbox{I\hspace{-0.82em}R}}

}{
\newcommand{\C}{{\Bbb C}} 
\newcommand{\I}{{\Bbb I}} 
\newcommand{\R}{{\Bbb R}} 
}
\makeatother

\newcommand{\De}{{\partial\mkern-9mu/}} 

\newcommand{\D}{i \De}
\newcommand{\A}{A\mkern-11.5mu/\,} 

\newcommand{\complex}{\C}

\makeatletter
\let\storepage=\thepage 
\def\thepage{\ifnum\c@page=1 \@empty \else \storepage \fi}
\makeatother
\newcommand{\pptnumbers}{\hfill \protect\parbox[t]{8pc}
 {\rm\hfill DSF--18/97\\
\null\hfill OUTP--97--17--P\\
\null\hfill hep-th/9704184}}

\title{\bf Mirror Fermions in Noncommutative Geometry}

\author{F.~Lizzi,$^{1,2,3}$ G.~Mangano $^{1,2}$, G.~Miele$^{1,2}$, 
G.~Sparano$^{2,4}$\thanks{%
E-mail:  \texttt{lizzi@na.infn.it}, \texttt{mangano@na.infn.it}, 
\texttt{miele@na.infn.it}, 
\texttt{sparano@na.infn.it}}\\[4pt]
$^1$Dipartimento di Scienze Fisiche, Universit\`a di Napoli {\it Federico
II}\\[2pt]
$^2$ INFN, Sezione di Napoli, Italy\\[2pt]
$^3$Theoretical Physics, 1 Keble Road, Oxford OX1 3NP, UK\\[2pt]
$^4$Dipartimento di Ingegneria Informatica e 
Matematica Applicata\\[2pt] 
Universit\`a di Salerno}

\begin{document}


\maketitle
\thispagestyle{myheadings} 
\markright{\pptnumbers} 

\begin{abstract}
In a recent paper we pointed out the presence of extra fermionic
degrees of freedom in a chiral gauge theory based on Connes
Noncommutative Geometry. Here we propose a mechanism which provides a
high mass to these mirror states, so that they decouple from low
energy physics. 
\end{abstract}
\newpage

Recently in Ref.\cite{Hilbert} we have pointed out a doubling of the fermionic
degrees of freedom in the Connes \cite{ConnesLott,RealNCG} approach to
chiral gauge theories based on Noncommutative Geometry
\cite{Book}. In this brief report we suggest a mechanism which solves
the problems posed by these extra degrees of freedom, by giving them
a high mass. The mechanism on which the solution is based is loosely
modelled in analogy with lattice gauge theories where a similar
phenomenon occurs. The origin of the two phenomena are however quite
distinct, at least with the present level of understanding. 

As is known the essential ingredients of Noncommutative Geometry are
an algebra ${\cal A}$, which encodes the topology of space--time (or
its noncommutative generalization), an Hilbert space ${\cal H}$ which
represents the fermionic mass degrees of freedom, and an operator $D$
which generalizes the Dirac operator and which encodes the metric
structure of the space. In addition to these three components of the
so called spectral triple, there are other two essential elements: the
real structure $J$, which represents charge conjugation, and
a grading $\gamma$ which generalizes the usual $\gamma_5$. 

At present the noncommutative geometric structure of chiral gauge theories is
understood as the product of continuous geometry representing the
usual (commutative) space--time times an internal geometry. Thus the
algebra is chosen as ${\cal A}=C^{\infty}(\R^4,\complex)\otimes{\cal
A_F}$, where $C^{\infty}(\R^4,\complex)$ is the algebra of smooth
complex valued functions on $\R^4$, and ${\cal A_F}$ a matrix algebra
whose unimodular group is the gauge group. Analogously ${\cal
H}=L^2(S_{\R^4}) \otimes {\cal H}_F$, with $L^2(S_{\R^4})$ the space
of spinors and ${\cal H}_F$ an internal Hilbert space which comprises
all fermionic degrees of freedom.  The other ingredients are obtained
in a similar way, for details we refer to the literature on the
subject in its various versions
(\cite{RealNCG,JoePepe}--\cite{ticos} and references
therein). Other versions of gauge theory based on Noncommutative
Geometry \cite{Coquereaux} have some of the basic ingredients of the
construction which differ in essential ways from the ones treated
here, and in general the considerations about mirror fermions will not
apply. 

It is evident that the full power of noncommutative geometry is still
used in a very limited way. The theory is some sort of Kaluza--Klein
in which there is a continuous commutative space time, still made of
usual points with the usual Hausdorff topology. At each point then
there is a noncommutative space of the simplest kind possible, the one
represented by finite dimensional matrix algebras. Despite the
promising phenomenological features of the model
\cite{ChamsConnes,IKS,MarsSpectr}, this simple choice of the space as a product
creates some problems. The main one arises in ${\cal H}$. For the
consistency of the model it must be the tensor product of {\em
spinors} times {\em all} fermionic degrees of freedom, and therefore
some degrees of freedom will appear more than once. Moreover the
chirality assignments of the extra degrees of freedom are incorrect.
We will be more detailed in the following. 

The problem could be solved by projecting out the unwanted degrees of
freedom, but as we showed in Ref.\cite{Hilbert} this procedure is
ambiguous and can only be made in a highly {\em ad hoc} fashion. 

In this paper we would like to explore another possibility, namely
that the mirror fermionic degrees of freedom are actually real ones,
but that the mass they have is too large to be detected, or to have
any effect at present energies. There might be some consequences for
the early universe, and we will comment on this later. 

In what follows we will work in Minkowski space. It has been already
stressed in Ref.\cite{Hilbert} that the appearance of mirror fermions is
independent of the choice of the signature. However, as will be clear
in the following, this aspect is crucial for the solution of the
problem we propose here. For the bosonic terms in the action, the
choice of the scalar product is not so important, since the euclidean
and lorentzian theories can be related by Wick rotation. This is not
true for fermionic terms. In this case, in fact, since the involved
representations are complex and the invariants are written in terms of
a hermitian form rather than a scalar product, there are no
transformations which can relate the positive definite hermitian form
of euclidean theories, with the ones, with no definite sign, of
lorentzian models. In other words, as far as the bosons are concerned,
the invariants under $SO(3,1)$ build up in terms of scalar product of
the fields become, via Wick rotation, the corresponding invariants
under $SO(4)$, whereas this does not occur for spinors. Note that, the
{\it euclidean} theories introduced as a way to regularize functional
integration are just the Wick rotation of the lorentzian models, and
thus not invariant, in general, under $SO(4)$. 

To explain the problem and the solution we propose, the subtleties of 
the full construction are unnecessary. We will therefore deal with a 
simplified model, in which a single generation contains only one spinor.
The generalization is absolutely straightforward.

As usual we start defining the various elements of the Connes
construction. For the example under consideration of a spontaneously
broken $U(1)$ theory, we start with the algebra ${\cal A}=
C^{\infty}(\R^4,\complex) \otimes \left( \complex \oplus \complex \right)$.
The unimodularity condition
will reduce the gauge group $U(1)_L \otimes U(1)_R$, the unitary
elements of the algebra, to $U(1)_A$. The Hilbert space has the usual
tensor product structure ${\cal H} = L^2(S_{\R^4}) \otimes {\cal H}_F$,
where  ${\cal H}_F = \complex^4$. A generic element of ${\cal H}$, can
be expressed as a linear combination of elements of the form
$\Psi=\psi \otimes h_F$, with $\psi$ a Dirac spinor in $L^2(S_{\R^4})$ and 
$h_F = \left(h_L, h_R, h^c_R,h^c_L \right) \in {\cal H}_F$.\\
On ${\cal H}$, an element $\alpha$ of ${\cal A}$ is represented 
as follows
\be
\rho(\alpha) =\left(\begin{array}{cccc} a_L(x) & & & \\
& a_R(x) & & \\ & & a^*_L(x) & \\ & & & a_R^*(x) \end{array}
\right)~~~,
\label{5}
\ee
where $a_L(x)$ and $a_R(x)$ belong to $C^{\infty}(\R^4,\complex)$.
By observing that for each $x \in \R^4$ a generic spinor $\psi$ can be 
decomposed as $\psi(x) = \psi_L + \psi_R + \psi_R^c + \psi_L^c$, 
among the 16 possible combinations in $\Psi=\psi \otimes h_F$, 
the following have a chirality {\it mismatch}
\be
(\psi_L+\psi_R^c)  \otimes (h_R + h_L^c) + (\psi_R+\psi_L^c)
\otimes (h_L+h^c_R)
~~~.\label{9}
\ee
These are the 
spurious degrees of freedom which behave as the mirror fermions on lattice 
gauge theories and should be eliminated from the theory.

As Dirac operator we consider the following generalization of the customary one
\be
D = \D \otimes \I + \I \otimes {\cal M} - \gamma_5 \otimes 
\gamma_F {\cal M}'~~~,
\label{10}
\ee
where $\gamma_F$ is the grading in the finite space ${\cal H}_F$
\be
\gamma_F = \left(\begin{array}{cc}\sigma_3 & 0 \\ 0 & - \sigma_3
\end{array}\right)~~~.
\label{11}
\ee
The grading in ${\cal H}$, denoted with $\gamma= \gamma_5 \otimes \gamma_F$
satisfies the following relations:$\gamma = \gamma^{\dagger}$,
$\gamma^2=\I$, and $\left\{\gamma,D\right\}=0$.

It is worth observing that $\gamma$ has eigenvalues $+1$ and $-1$ on
the physical and unphysical fermionic states, respectively, and
therefore the combination $P = { (\I + \gamma)/ 2}$ is the projection
operator on the subspace of physical states. Interestingly one of the
natural structure in the algebraic construction of gauge theories in
Noncommutative Geometry, namely the grading $\gamma$, distinguishes
the real physical fermionic states in the underlying Hilbert space
from their mirror partners. 

At this point one could consider that the solution of the problem is
to project out the unwanted degrees of freedom with the {\em natural}
operator $P$. However, we showed in Ref.\cite{Hilbert} that this projection
must be done only in the fermionic part of the action, since for the
bosonic part it would eliminate the self--dual or the anti self--dual
part of the gauge tensor fields. One can still be satisfied with
treating the two terms of the action in a different way, but, apart
from issues of naturality and aesthetics, this leaves open the problem
of definition of the actual Hilbert space of the theory. 

In equation (\ref{10}) ${\cal M}$ and ${\cal M}'$ are the mass matrices,
defined as usual by
\be
{\cal M}=\left(\begin{array}{cccc} 0 & m & 0 & 0\\
m^* & 0 & 0 & 0 \\ 0 & 0 & 0 & m^*\\
0 & 0 & m & 0\end{array}\right)~~~,
\label{14}
\ee
and similarly for ${\cal M}'$. In this simple example $m$ and $m'$ are just
numbers, while in more complex cases, like the Standard Model, they would be 
matrices.

Finally, the real structure $J= J_D \otimes J_F \cal{C}$,
${\cal C}$ being complex conjugation, satisfies the relations
\begin{eqnarray}
[J,D]=0  ~~~,~~~~J \gamma =  \gamma J~~~,~~~~~
[\alpha,J^{\dagger} \alpha' J]=[[D,\alpha], J^{\dagger} \alpha' J]=0
~~~,~~~~\alpha,\alpha' \in {\cal A}~~~. \label{16}
\end{eqnarray}
Using these conditions one easily find that $J_D= \gamma_2$ and
\be
J_F = \left(\begin{array}{cc} 0 & \I \\
\I & 0 \end{array}\right)~~~.
\label{17}
\ee
All relations uniquely fix $J$ up to a phase and are compatible with the
generalized Dirac operator introduced in (\ref{10}) with a new mass matrix
${\cal M}'$.

It is important to notice at this point that this is not true anymore
if we try to implement the same structure in euclidean space. In fact,
in this case, starting with the same algebra, grading and Hilbert
space, the euclidean Dirac operator has the form 
\be
D = \De \otimes \I + \I \otimes {\cal M} - 
\gamma_5 \otimes \gamma_F {\cal M}'~~~,
\label{18}
\ee
where, for example, we choose all $\gamma^{\mu}$ to be antihermitean.

The condition $[D,J]=0$ in particular requires the following conditions,
as it is easy to verify
\be 
J_D \gamma_{\mu}^* = \gamma_{\mu} J_D~~~,~~~~ [J_F {\cal C}, {\cal M}]=0
~~~,~~~~[J, \gamma_5 \otimes \gamma_F {\cal M'}] 
= 0~~~.
\label{20}
\ee
The first two give, up to a phase, $J_D=\gamma_0 \gamma_2$, and $J_F$
as in Eq.(\ref{17}). Thus the third relation of Eq.(\ref{20})
implies $m' \{\gamma_5,J_D\}=0$ which is satisfied only if ${\cal M}'$ 
identically vanishes. 

This is also an example which shows that working in euclidean space, while 
having advantages for the definitions of the mathematical objects involved 
in the theory, can have some non--trivial consequences from the physical 
point of view.

Going back to the Minkowski case, in order to construct the lagrangian
density for the abelian model we are considering, we first need to
construct the gauge connection one-forms, as the elements
$\rho=\alpha[D,\alpha']$. Removing junk forms we get 
\be 
\rho= \left( \begin{array}{cc} \A_L & (\phi -\phi_0) - \gamma_5 (\phi'
-\phi'_0) \\
(\phi^\dagger - \phi_0^{\dagger}) - \gamma_5 (\phi^{'\dagger} - 
\phi_0^{' \dagger})
& \A_R \end{array} \right)~~~,
\label{21}
\ee
where the Yang--Mills connection (one--form) 
$\A_{L,R} \equiv \sum_i a^{'i}_{L,R} \D a^i_{L,R}$, with the condition 
$A^{\mu}_{L,R} = A^{\mu*}_{L,R}$
and with $\phi - \phi_0 \equiv \sum_i a^{'i}_L({\cal M} 
a'_R - a'_L {\cal M})$.
A similar expression holds for $\phi'-\phi'_0$. Unimodularity
condition $Tr(\rho)=0$ reduces, as already mentioned, the gauge group
to the axial term only $U(1)_A$, tracing out the vector part 
\be 
\rho= \left( \begin{array}{cc} \A & (\phi -\phi_0) - \gamma_5 (\phi'
-\phi'_0) \\
(\phi^\dagger - \phi_0^{\dagger}) - \gamma_5 (\phi^{'\dagger} - 
\phi_0^{' \dagger})
& -\A\end{array} \right)~~~.
\label{22}
\ee
The two Higgs fields $\phi$ and $\phi'$ represent the connection
fields in the discrete direction and are related to the terms
proportional to ${\cal M}$ and ${\cal M}'$, respectively. Under a
gauge transformation, represented by the unitary elements $u$ of the
algebra ${\cal A}$, with the condition $u_L=u_R^*$, $\rho$ transforms
as $\rho \rightarrow u[D,u^*] + u \rho u^*$.

By using the matrix representation for the algebra and Eq.(\ref{22}), we
see in particular that both $\phi$ and $\phi'$ have equal non
vanishing charge with respect to the $U(1)_A$ gauge group. Their
expectation values on the vacuum state would therefore break axial
gauge symmetry. 

The bosonic contribution to the action of the model can be obtained by
evaluating the square of the curvature $\theta=d \rho + \rho^2$,
traced over the entire fermion Hilbert space. 
\be
S_B = Tr \theta^2 = \int d^4 x \left[ -{1 \over 4} F_{\mu \nu} 
F^{\mu \nu} +
(D_{\mu} \phi) (D^{\mu} \phi)^* + (D_{\mu} \phi') (D^{\mu} \phi')^*
-V(\phi,\phi') \right]~,
\label{24}
\ee
where $F^{\mu \nu}$ is the tensor field for the axial gauge 
potential $A^{\mu}$, and $D_{\mu}= i\partial_{\mu} + 2 A_{\mu}$.
The Higgs potential $V(\phi,\phi')$ takes the form
\be
V(\phi, \phi') \propto ( |\phi|^2 + |\phi'|^2 - \mu^2)^2 +
(\phi \phi^{'*} + \phi^* \phi' - \lambda^2)^2~~~,
\label{25}
\ee
where $\mu^2=|\phi_0|^2 + |\phi'_0|^2$ and 
$\lambda^2= \phi_0 \phi_0^{'*} + \phi_0^* \phi'_0$.

Finally, the fermionic action is expressed in terms of the invariant scalar
product
\be
S_F = \langle \Psi,(D + \rho + J^{\dagger} \rho J ) \Psi \rangle ~~~,
\label{26}
\ee
which gives, together with kinetic and interaction terms with gauge 
potential and Higgs fields, the following mass terms
\bea
S_F \mbox{(mass terms)} = \int d^4 x 
(\bar{\psi} \psi) ~(h_F^\dagger {\cal M} h_F) -
(\bar\psi \gamma_5 \psi)~ (h_F^\dagger \gamma_F{\cal M}' h_F) \nonumber\\
= \int d^4 x \overline\Psi
\left( \left({1+\gamma\over 2}\right)
\left({\cal M}-{\cal M}'\right)+
\left({1-\gamma\over 2}\right)
\left({\cal M}+{\cal M}'\right)\right)\Psi~,
\label{27}
\eea
where $\psi \in L^2 (S_M)$ and $h_F \in {\cal H}_F$. Decomposing $\psi$
and $h_F$ as shown before it follows that all fermions
belonging to the physical subspace of ${\cal H}$ acquire mass equal to
$m-m'$, while their mirror partner get instead $m+m'$. In particular,
if both $m$ and $m'$ are very large, namely if the breaking of
$U(1)_A$ occurs at a very high scale, all mirror states completely
decouple from the low energy theory. If all physical states should
remain instead massless or take a very small mass term, one has to
impose $m-m' <\!< m,m'$. In this scheme this fine-tuning condition
seems to be unavoidable. 

This mechanism which gives a high mass to the mirror fermions via a
spontaneous breaking at a high scale, may provide a scalar dynamics
which would drive chaotic inflation. We have already discussed the
appearance of inflation in noncommutative geometry models in
Ref. \cite{Napolinfla}. 

There is only a drawback in this solution of the doubling problem. To
obtain masses which are very high for mirror states only, and 
practically vanishing for the observed
fermions, one has a
serious problem of fine tuning. Actually, fine tuning problems are not new to
Noncommutative Geometry \cite{ChamsConnes}.

Despite this feature, the mechanism we propose here is a
dynamical way of eliminating the problem. Actually, we think that the main
point is the fact that we are too naive in considering the
geometry as the product of a continuous commutative space, times the space
of finite dimensional matrix algebra. The real structure of space time is
probably a more complicated one, and Noncommutative Geometry seems the ideal
tool to study its structure. In this respect, it is worth pointing out, to
conclude, that the solution of mirror fermion problem we have discussed
is suggesting that Planck mass or some other high mass scale should be the
natural scale where noncommutative structure of space time should manifest
itself. Shadows of these effects, as for example a particular
tensor product form for the fermion Hilbert space, or the choice for the
algebra, could well be present for low energy theories as the Standard Model.

\subsection*{Acknowledgments}

\medskip

FL would like to thank the department of Theoretical Physics of the
University of Oxford for hospitality.
\bigskip

\def\up#1{\leavevmode \raise.16ex\hbox{#1}}
\newcommand{\npb}[3]{{\sl Nucl. Phys. }{\bf B#1} \up(19#2\up) #3}
\newcommand{\plb}[3]{{\sl Phys. Lett. }{\bf #1B} \up(19#2\up) #3}
\newcommand{\revmp}[3]{{\sl Rev. Mod. Phys. }{\bf #1} \up(19#2\up) #3}
\newcommand{\sovj}[3]{{\sl Sov. J. Nucl. Phys. }{\bf #1} \up(19#2\up) #3}
\newcommand{\jetp}[3]{{\sl Sov. Phys. JETP }{\bf #1} \up(19#2\up) #3}
\newcommand{\rmp}[3]{{\sl Rev. Mod. Phys. }{\bf #1} \up(19#2\up) #3}
\newcommand{\prd}[3]{{\sl Phys. Rev. }{\bf D#1} \up(19#2\up) #3}
\newcommand{\ijmpa}[3]{{\sl Int. J. Mod. Phys. }{\bf A#1} \up(19#2\up) #3}
\newcommand{\prl}[3]{{\sl Phys. Rev. Lett. }{\bf #1} \up(19#2\up) #3}
\newcommand{\physrep}[3]{{\sl Phys. Rep. }{\bf #1} \up(19#2\up) #3}
\newcommand{\jgp}[3]{{\sl J. Geom. Phys. }{\bf #1} \up(19#2\up) #3}
\newcommand{\journal}[4]{{\sl #1 }{\bf #2} \up(19#3\up) #4}

\end{document}